\documentstyle{article}
\def\P3M{\hbox{$P^{3}M$}}
\def\PP{\hbox{$PP$}}
\def\PM{\hbox{$PM$}}

\def\AP3M{\hbox{$AdP^{3}M$}}

\def\cc2{c2}
\def\cc3{c3}
\def\cc4{c4}
\def\cc{c}

\def\AaA{A\&A}

\def\ApJ{ApJ}
\def\ApJL{ApJ}
\def\ApJS{ApJS}

\def\ARAA{ARA\&A}

\def\JCP{J.~Comp.~Phys.}
\def\MN{MNRAS}
\def\Nat{Nat}

\def\PhD{PhD thesis}

\def\prep{in preparation}

\def\ref{\par \noindent \hang}
\def\etal{{\it et al.\thinspace}}
\def\eg{{\it e.g.\ }}

\def\spose#1{\hbox to 0pt{#1\hss}}
\def\approxlt{\mathrel{\spose{\lower 3pt\hbox{$\sim$}}
	\raise 2.0pt\hbox{$<$}}}
\def\approxgt{\mathrel{\spose{\lower 3pt\hbox{$\sim$}}
	\raise 2.0pt\hbox{$>$}}}

\def\<{\thinspace}
\def\boxit#1{\vbox{\hrule\hbox{\vrule\kern3pt\vbox{\kern3pt
          #1 \kern3pt}\kern3pt\vrule}\hrule}}

\def\ga{{\rm\thinspace gauss}}

\def\h50{\hbox{$\rm\thinspace h_{50}$}}
\def\h50m1{\hbox{$\rm\thinspace h_{50}^{-1}$}}
\input epsf
\begin{document}
\huge
\centerline{\bf Hydra: A Parallel Adaptive Grid Code}
\vspace{1 cm}
\Large
\centerline{
F.R.Pearce\footnote{Physics Department, University of Durham, Durham, UK}$^,$\footnote{Astronomy Centre, University of Sussex, Falmer, Brighton, UK},
H.M.P.Couchman\footnote{Department of Physics and Astronomy, 
University of Western Ontario, Ontario, Canada}
}
\vspace{0.5cm}
\centerline{F.R.Pearce@durham.ac.uk, couchman@coho.astro.uwo.ca}
\vspace{1cm}
\large

\noindent{\it Subject headings:} methods: numerical -- cosmology: theory -- hydrodynamical simulation

\section*{Abstract}

We describe the first parallel implementation of an adaptive 
particle-particle, particle-mesh code with smoothed particle
hydrodynamics. Parallelisation of the serial code, ``Hydra'', is
achieved by using CRAFT, 
a Cray proprietary language which allows rapid implementation
of a serial code on a parallel machine by allowing global
addressing of distributed memory. 

The collisionless variant of the code has already completed several
16.8 million particle cosmological simulations on a 128 processor Cray
T3D whilst the full hydrodynamic code has completed several 4.2
million particle combined gas and dark matter runs. The efficiency
of the code now allows parameter-space explorations to be performed
routinely using $64^3$ particles of each species. A complete run
including gas cooling, from high redshift to the present epoch
requires approximately 10 hours on 64 processors.

In this paper we present implementation details and results of the
performance and scalability of the CRAFT version of Hydra under
varying degrees of particle clustering.

\section{Introduction}

A key goal of contemporary cosmology is to understand the growth of
structure; that is to connect the spectrum of linear fluctuations
which exists at the epoch of recombination (at present unknown) with
the highly non-linear galaxies and clusters that we observe at
present. Simulations of cosmic structure are amongst the
most ambitious supercomputer applications: the huge range in observed
structures, from sub-galactic to the largest superclusters and voids,
perhaps a factor of $10^9$ in mass, presents a formidable numerical
challenge.

The large range of density contrasts and complicated geometries that
arise in cosmological simulations make Lagrangian particle methods
popular choices for modelling the growth of structure. Typically, we
wish to model a representative volume of the universe (corresponding
to $\sim0.1\%$ of the observable volume) over a period of 10 billion
years. A contemporary simulation might use of order 10 million
particles to represent the matter density in the universe. Initially
these particles are distributed nearly uniformly within a triply
periodic cube. The simulations follow the motion of these particles
as they move under the influence of self gravity (and
short range gas forces if appropriate). The push for ever higher
resolution in such simulations has inevitably led to the use of
parallel supercomputers, both because of the total processing power
available but also, equally importantly, in order to satisfy the huge
memory demands of these programs.

The trend towards the use of parallel supercomputers for large-scale
cosmological simulations is clear, but is not without problems. We will
address many of the issues confronting parallel particle codes in this
paper, with emphasis on those of particular importance to
particle--grid codes. In general, Lagrangian
particle methods are hard to load balance. Static load balancing
techniques do not work well because the distribution of particles
relative to the necessary underlying Eulerian computational framework
changes dynamically, and can become highly non-uniform. Further, the
locations at which structures develop are not known accurately
beforehand.  This causes severe problems even for dynamic load
balancing strategies. In this paper we outline our approach to solving
these problems through the use of an adaptive grid refinement
technique. This reduces the total amount of computational work and
allows us to break the problem into many smaller parts which can then
be efficiently distributed over the processors. We conclude this part
of the introduction with a brief overview of the development of N-body
particle techniques. More details can be found in the reviews of
Sellwood~(1987) and Couchman~(1997).

Since the pioneering work of von Hoerner (1960) and Aarseth (1963)
significant advances have been made towards numerical solution of the
gravitational N-body problem. The early codes used a direct summation
(here ``particle-particle'' or PP) approach, directly accumulating the
force on each particle from the contributions of all
other particles.  The computing time required for this method scales
as $O(N^2)$ and for more than a few thousand particles becomes
prohibitively expensive, unless application-specific hardware is
used (\eg GRAPE boards, see Steinmetz 1996).

The limit on particle number imposed by the undesirable scaling
properties of PP codes has been circumvented in more recent codes by
approximating the gravitational potential. All of
these methods correspond in some way to a decomposition of the field
due to distant particles in terms of various 
basis functions. Two general
approaches have been popular.  The method used first (and from which
the method we describe in this paper is derived) was to represent the
gravitational field on a mesh of 
fixed resolution (e.g., Miller 1978, Miller \& Smith 1980). Efficient
solution methods for Poisson's equation on regular meshes, such as the
FFT convolution, allow 
an algorithm with computation time which scales as $\sim O(N\log N)$,
although the force resolution is limited to roughly two grid
separations. These types of methods will be referred to here as
Particle--Mesh (PM) codes.

More recently codes have been developed which represent the field via
a multipole expansion.  The two main variants of this approach are the
various ``tree codes'' (Appel 1985, Jernigan 1985, Barnes \& Hut
1986), and the true multipole expansion methods (van Albada \& van
Gorkum~1977, Villumsen~1982, Greengard \& Rokhlin~1987). These methods
have an advantage over grid-based codes in that their resolution is
not limited by the grid scale. This is desirable in order to be able
to correctly model bound objects at fixed physical separation within
a simulation cube which is comoving with the universal expansion. It is
possible to use a very fine mesh in order to reproduce forces over a
wider range of scales but this increases both the storage requirement
and computation time, as well as leading to inefficiency in regions of
low particle density. An alternative method for overcoming the lack of
sub-grid scale forces in conventional grid schemes is to combine these
techniques with PP methods.  In these hybrid codes the direct sum is
performed only over nearby neighbours out to a distance necessary to
augment the fixed resolution grid force to the force required overall. 

Grid-based codes, in particular particle-particle, particle-mesh,
(P$^3$M) codes, have been applied to a variety of problems, ranging
from the formation of large scale structure and galaxy clusters to
galaxy mergers 
and galaxy formation (e.g., Efstathiou \etal 1985, 1988, Pearce, Thomas \&
Couchman 1993, 1994).  Several authors (Evrard 1990, Thomas \&
Couchman 1992) have incorporated Smoothed Particle Hydrodynamics
(SPH---Monaghan 1992) to model gas processes. P$^3$M is very efficient in
terms of memory use, but in general is very expensive when the amount of
clustering becomes high and the number of neighbours in the PP sum
increases. The use of adaptive refinement to increase the grid
resolution in selected regions (Couchman~1991) has ameliorated this
problem and the latest workstations can run simulations with $N \sim
7$ million particles in around a month (Cole \& Lacey 1996).

The P$^3$M method does not lend itself easily to parallelisation for a
number of reasons. As described above, the mapping from Lagrangian
particle space to the Eulerian grid can lead to load balancing
problems and difficulties with data decomposition on distributed
memory computers. The advantages of grid-based codes in terms of
memory usage and the computational simplicity of regular grids are
substantially reduced in comparison to the more general data-structure
employed by ``Tree'' codes by these parallelization issues. Efficient
parallel PM implementations have been constructed (Ferrell \&
Bertschinger  
1994), and there are parallel PP schemes (Ferrell \& Bertschinger
1995), it is difficult, however, to produce a combined method that makes
efficient use of memory and has a good communication strategy.  We
discuss the parallelisation of a grid-based code here; the
parallelisation of treecodes has been discussed elsewhere, for
example, Dubinski (1996), Salmon (1991), Dav\'e, Dubinski
\& Hernquist (1997).

\smallskip
The British supercomputing community has invested in a Cray T3D which
is installed in Edinburgh. This is a massively parallel, distributed
memory MIMD (multiple instruction, multiple data) machine with 512
processors. These processors are linked together in a toroidal network
with high speed interconnections designed to minimise the overhead of
inter-processor communication. Each node has a 150Mhz DEC Alpha EV4
processor and associated custom hardware, and has 64Mb of memory per
node. 

In this paper we will describe a parallel implementation of a serial
P$^3$M--SPH code, ``Hydra'' (Couchman, Thomas \& Pearce~1995), for the
T3D platform using Cray's proprietary CRAFT software. This code
(parallel Hydra) is the first parallel adaptive particle-particle,
particle-mesh plus smoothed particle hydrodynamics algorithm. CRAFT
has enabled us to quickly modify the serial algorithm for parallel
execution.

The paper is laid as follows. In section 2 we describe the serial
operation of Hydra and in section 3 deal with the problems of porting
the code to parallel machines.  In section 4 we look at the code's
performance on the Cray T3D, in particular the load balance and speed
for both clustered and unclustered particle distributions.

\def\ga{\mathrel{\hbox{\rlap{\hbox{\lower4pt\hbox{$\sim$}}}\hbox{$>$}}}}

\section{Hydra}

Hydra---an implementation of Smoothed Particle Hydrodynamics (SPH)
in an adaptive P$^3$M code (Couchman, Thomas \& Pearce 1995) is the
endpoint of several stages of code development which we describe below
before giving a more detailed description of Hydra itself.

The fundamental drawback of the PP method, as noted above, is the
rapidly increasing computational cost of the $O(N^2)$ sum over all
particle pairs, which becomes prohibitive for $N$ greater than a few
thousand.  The most straightforward way of circumventing this problem
(borrowed initially from plasma physics) involves
smoothing the particles onto a uniform grid. The potential
corresponding to the sampled mass distribution is solved for 
using a fast Fourier transform (FFT) convolution method. The grid potential
is then differenced and the forces interpolated back to the particle
positions.  The computational cost of the ``particle-mesh'' ($PM$)
method scales as $\alpha N + \beta L^3\log L$, where $\alpha$ and
$\beta$ are constants, $N$ is the number of particles and $L$ is the
size of the mesh.  Typically $N \sim (L/2)^3$, which suggests the 
familiar $\sim O(N \log N)$ scaling.

The main drawback of this approach is that it cannot reproduce
structure on scales less than roughly 2 mesh spacings; the Nyquist
wavelength of the grid.  In three dimensions improving the resolution
by increasing the grid size is problematic due to the rapidly
increasing memory requirement: a single grid with $512^3$ cells
requires $128\,$Mwords of memory.  The advantages of FFT-based PM schemes
are that they are extremely fast, the speed is independent of the
degree of particle clustering and the boundaries are automatically
periodic. The latter feature, a by-product of using an FFT potential
solver, is a useful means of modelling a section of a very much
larger, or infinite, universe.

The next advance, deriving again from plasma physics (Hockney \&
Eastwood 1981), was to combine the PP and PM methods to
produce the hybrid ``particle-particle, particle-mesh'' (P$^3$M)
scheme. This method was first used in Astrophysics by Efstathiou 
\& Eastwood~(1981). This method splits the gravitational force into two
components, long- and short-range. The PM method is used to solve for
the long-range component and, on scales smaller than about two grid
spacings the `soft' mesh force is augmented by a direct PP sum over near
neighbours. The calculation of the PP part proceeds by binning
particles onto a coarse mesh which is then used to search efficiently
for near neighbours to include in the short range component of the
force calculation.

The cputime for this method scales as;
\begin{equation}
 t_{\rm P^3M} \sim  \alpha N + \beta L^3\log L + \gamma\sum_{pp^\prime} N_{p}N_{p^\prime}
\end{equation}
where the first two terms come from the PM scaling from above and
$\gamma$ is a constant. $N_p$ is the number of particles in the coarse
PP mesh cell $p$; the sum over $p$ is over all cells in the mesh,
whilst $p^\prime$ ranges over the cell $p$ and its 26 neighbour cells.  If the
particles are uniformly distributed throughout the computational
volume and the PM mesh has been chosen such that $L^3\ga N$, then
$N_{p}$ will be a small constant and the scheme scales roughly as $O(N
\log N)$ as with the PM method.  This technique produces a scheme with much
better spatial resolution than the PM method and much greater speed
than the PP method. The difficulty occurs when heavy clustering is
present as then the number of neighbours, $\sum N_{p^\prime}$ is
spatially variable and may become very large. Under these conditions
the $O(N^2)$ nature of the third term of equation~(1) begins to
dominate and the overall runtime increases rapidly.

This problem was solved by automatically placing higher resolution
sections of grid within 
the original framework of the standard P$^3$M scheme
(Couchman 1991). In this case computational ``hotspots'' are
identified where the work of the short-range PP calculation will be
high and a finer grid is placed in this region. This reduces the range
of the PP interaction and results in a corresponding
decrease in $\sum N_{p^\prime}$. This lowers the overall computational cost
despite an 
increase in the grid-based work of the first and second terms
in equation~1 above.

Finally, smoothed particle hydrodynamics (SPH) was incorporated into
the adaptive P$^3$M code to form Hydra. SPH is a hydrodynamic
technique in which thermodynamic quantities are carried by
particles. Values of these quantities which are required at any point
in the fluid are interpolated from neighbouring particles. Since the
interactions are short-range, SPH forces are easily computed within
the same framework that performs the PP component of the gravitational
force. 

In practical versions of these algorithms simple timestepping schemes
such as leapfrog or low order predictor--corrector-type schemes are
used. This is primarily because the cost of the force calculation
prohibits many force evaluations per step and storage of quantities
for each of $N$ particles must be minimized.

Full details of the techniques used to construct the Hydra code,
including consideration of various time-stepping schemes, are
described by Couchman, Thomas \& Pearce~(1995). The serial version of
this code is available to the community from either of the
following web sites:

\begin{itemize}
\item http://coho.astro.uwo.ca/pub/hydra/hydra.html

\item http://star-www.maps.susx.ac.uk/\~\/pat/hydra/hydra.html
\end{itemize}

In the following sub-sections we discuss the three key features of the
algorithm; force matching, the organization of particles
onto grids and the strategy for placing refinements. 

\subsection{Force matching}

The central idea behind the P$^3$M algorithm is the explicit splitting
of the force into a long-range component calculated by a PM technique
and a short-range component accumulated by the direct PP sum. Accurate
forces demand that these two components be carefully matched. A very
significant advantage accrues from using a Fourier technique to
calculate the PM force: one is free to adjust the components of the
Green's function in Fourier space to produce any force shape desired
(within the limits imposed by the harmonic content that can be handled
by the finite Fourier transform). In particular, this permits a force
to be used which differs from $1/r^2$ only below a fixed radius (which
then becomes the PP search radius) and allows a significant reduction
in mesh errors, both fluctuation and anisotropy, to be achieved (Hockney
\& Eastwood~1981). The resulting pairwise grid force may be calculated
analytically.  The PP contribution is then calculated as the
difference between the grid force and the overall force required. As
shown in Figure~1, the long range force that results from the FFT is
the $r^{-2}$ of gravity but at short range it falls below this. The
smoother the grid force the smaller the mesh errors but the larger the
PP search radius must become to correctly augment the grid force to
the required $r^{-2}$.  Hydra sets the PP search length automatically
to keep the error below a specified amount.  For an input maximum
error of 7.7 percent, Hydra searches out for PP neighbours up to 2.16
times the grid spacing, leaving a mean residual error of around 0.5
percent.

\begin{figure*}
 \centering
 \epsfysize=7cm\epsfbox{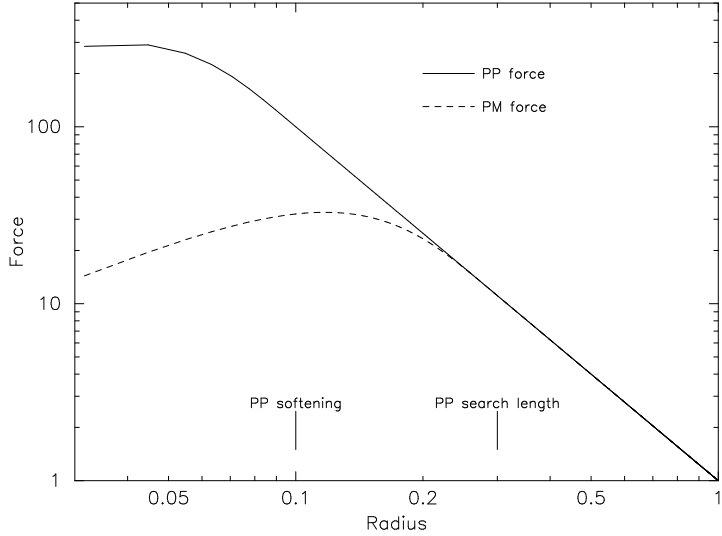}
 \caption{Matching the PP force onto the PM force. Hydra adjusts the
mesh softening to keep the error below the specified input error and
then chooses the PP search length automatically to match the forces.
}
 \label{force}
\end{figure*}

Refinements are handled in a similar fashion. The key idea is that the
force on a particle can now be the sum of several components, the PM,
as before, plus the force from a number of refined meshes (several
layers of refinements may be stacked recursively one within the other)
followed by an appropriate PP contribution.  Note that this force
accumulation strategy is quite different than in standard mesh-refinement
techniques in which the full force is derived directly from the 
potential on a refined grid. The final PP forces now have a much
shorter range (see Figure~2).
 
\begin{figure*}
 \centering
 \epsfysize=7cm\epsfbox{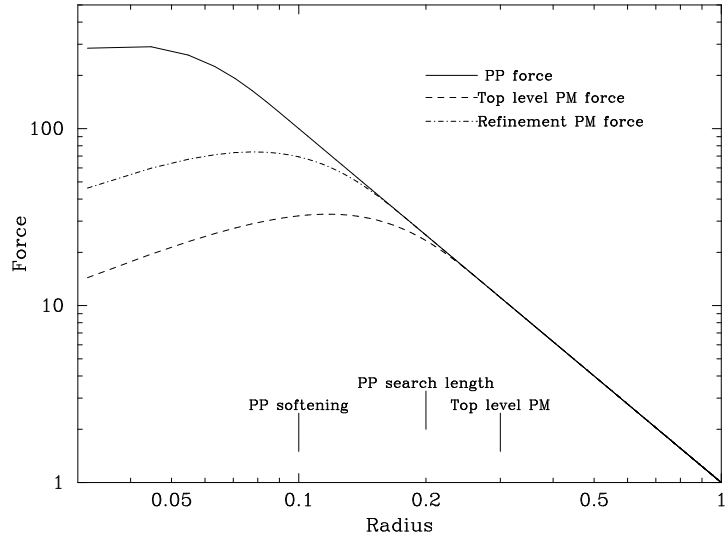}
 \caption{Matching the PP force onto the PM force with one
refinement. The PP search length (and hence PP work) has been reduced
by using a higher resolution mesh in this region.
}
 \label{refforce}
\end{figure*}

The errors incurred by using this technique to reconstruct the force
are small. In Figure~3 we show the root mean square difference in the
forces with and without refinement for a highly evolved position which
included up to 4 levels of refinement. As the parameter that controls
the accuracy of the force matching is reduced so is the mean
error. Particles outside refinements have exactly the same force
whether or not refinements are used.  Figure~3 also shows that the SPH
gas forces are calculated correctly across refinement boundaries. The
construction of the code ensures that the recovered {\em gas} pressure
forces are identical whether or not refinements are present.

\begin{figure*}
 \centering
 \epsfysize=7cm\epsfbox{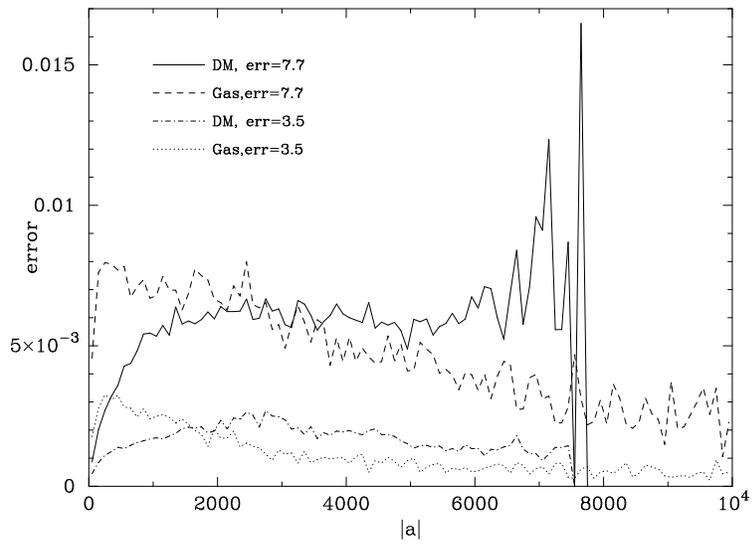}
 \caption{RMS force error versus acceleration for two different
error tolerances. For a tolerance of 7.7 the mean 
RMS error is 0.5 percent falling to approximately 0.2 percent for an
input tolerance of 3.5.
}
 \label{forceerror}
\end{figure*}

\subsection{Gridding schemes}

Hydra uses two independent gridding schemes to partition the particles
into local volumes. The first of these is the grid associated with the
FFT routine, the PM mesh. This is currently always a power-of-2 in
size because of the FFT method we use. The second is the PP cell grid,
over which neighbours are searched for by the PP routine.  The size of
this grid relative to the PM grid is determined automatically by Hydra
from the supplied error parameter as described above.  Refinement
placing is determined by the PP grid; the amount of PP work is
directly related to the number of neighbour particles in those
cells. One PP cell is typically 2.2 PM mesh separations across, so for
a $256^3$ mesh there are typically $116^3$ (around 1.6 million) PP
cells.

\subsection{Placing and computing refinements}

Gravitational clustering of the particles in the simulation volume
inevitably leads to an increase in the number of PP neighbours and
hence the cost of searching over these particles. This can lead to a
dramatic loss in efficiency. This serious flaw in the
standard P$^3$M algorithm has been alleviated in Hydra by replacing
regions with expensive PP sums by a finer mesh (or sequence of meshes)
followed by a PP sum with a much smaller search radius.

The decision as to which PP cells to refine is made after the PM
calculation but before the PP calculation. PP cells are refined
subject to the constraint that the refinements cover a whole number of
PP cells, are cubic and cannot overlap with other refinements. New
refinements are added to a list to be solved in sequence. Once
refinements have been placed, the PP calculation for the remaining
uncovered cells proceeds by summing over particle pairs in
neighbouring PP cells which do not {\it both} lie in the {\it same}
refinement. The effect of this procedure is that after this step 
only self forces are required on particles within a refinement to
complete the force. Refinements are calculated sequentially until none
remain by using a new P$^3$M calculation; first a PM cycle with
isolated boundary conditions and an appropriately shaped force,
followed by the possible addition of new refinements to the list and
finally a PP sum.

Refinements are placed to try to minimize the work in PP cells
containing many particles. This is achieved by finding peaks in the
distribution of particle counts in PP cells and placing refinements
there. These refinements are allowed to grow in size and move until a maximum
in the cost saving is achieved. A difficulty is that refinements
are not allowed to overlap. A small number of iterative sweeps are
made of the trial refinement positions so that they can settle into
the optimum distribution. A problem may occur when two clusters are
moving towards each other. When well separated they may each be
covered efficiently by separate refinements. However, because particle
pairs must be calculated between PP cells in {\em different}
refinements, there is a stage when it is more efficient to cover both
merging clusters by a singe refinement. To determine if this is the
case the distribution of counts in cells is smoothed on a large
scale---to artificially merge the peaks---and the refinement placement
strategy is repeated to determine if an increased saving can be
achieved. A final pass is performed to pick up any remaining cells
which may have been missed in the previous steps.
%The average number of particles in a refined cell is approximately
%10--30. 

Since all these calculations involve only integer operations on the
counts in PP cells, the procedure is fast and takes no more than a few
percent of the total cycle time for the serial code.

\section{Parallel Hydra}

Gravitational simulation codes require care to parallelise because of
the long range nature of the pair interaction. In the most
straightforward PP code each particle would need to receive
information about all others. This requirement makes parallelisation
difficult on distributed memory machines where the memory is broken up
into blocks each of which is associated with an individual processor.
Accessing 
data which resides in a different processor's memory requires
communication which is generally very slow compared with fetching
data from local memory. 

%Even on shared memory machines, care must be
%taken to use cache efficiently and not to saturate the path to memory.

At first sight it might appear that the P$^3$M algorithm provides a
perfect solution to this long range dilemma: the long range component
is cast into Fourier space in which it becomes fully local and the
short range part has a spatially limited extent. As will be shown
below this optimism is realised to a large extent for unclustered
distributions. However, as mentioned in the introduction, as
clustering develops, it becomes hard to efficiently distribute the
data across the processors and retain a good balance of computation
amongst the processors. There is also a more profound problem with
these codes in that the communication to computation cost tends to be
high; the particle is the fundamental unit and very little computation
is performed without reference to other particles. This requires that
we must always take care to organise the data efficiently to
minimize the communications overhead. This situation is in contrast to
that which holds with finite
element codes, for example, in which the element is the basic
unit and a large amount of calculation is performed within it without
reference to other elements.

To save time coding with explicit communication protocols we have used
CRAFT, a Cray directive-based language which permits the programmer to
address memory as a flat global address space.  CRAFT is similar to
HPF (High Performance Fortran), but with important extensions. It
essentially makes a distributed memory computer look like a shared
memory machine.  Under CRAFT remote memory addresses can be accessed
directly and the necessary communications are hidden, although a time
penalty is incurred. Care must still be taken in some aspects of data
distribution and program structure in order to obtain the best
performance. 

Hydra breaks down into 4 distinct parts, the long- and short-range
force calculations, building lists and placing the refinements. Each
of these tasks is modular in the serial code and this structure has
been retained in the parallel code. It is a measure of the utility of
CRAFT that we were able to transfer the serial code without major
reprogramming or restructuring of the code. This would not have been
the case had we employed an explicit message passing scheme. We look
at parallelising each of the four main blocks in turn.

\subsection{Long-range force calculation (PM)}

The long range force calculation involves smoothing the particle
masses onto a grid to obtain a sampled density distribution and then
performing a convolution to obtain the grid potential. The convolution
is efficiently calculated using Fast Fourier Transforms. Forces are
obtained in real space by differencing the grid potential. These
forces are then interpolated back onto the particles.  We can consider
each of these operations separately when we try to parallelise them.

Smoothing onto a grid is a well studied but difficult problem to
parallelise efficiently in clustered environments because many
processors may write to the same data (grid) location
simultaneously. It is also difficult to balance data storage and the
amount of work each processor has to do whilst simultaneously keeping
remote accesses to a minimum (Ferrell \& Bertschinger 1994, Pearce
\etal 1995).

The ability of CRAFT to make the distributed memory of a MIMD machine
look like shared memory makes the task of parallelising the code very
much 
easier. We employ a fixed distribution of the particles and grids
among the processors, so that each of the $N_{\rm pes}$ processors stores and
operates upon $N/N_{\rm pes}$ particles and holds $L/N_{\rm pes}$
sheets in the $z$ direction of
the PM mesh. 

Particles are distributed onto the mesh points using the
triangular shaped cloud kernel function (Hockney \& Eastwood 1981) by
using the {\em atomic update} facility of CRAFT.
The {\em atomic update} is a profoundly useful facility available
under CRAFT. It is a ``lock, fetch, increment, store, unlock''
directive that prevents a race condition occurring when two or more
processors try to write to the same memory location. Using this
directive many processors can simultaneously try to increment the same
grid point but must do so in sequence because each processor locks the
memory location whilst it is doing the update. The hardware lock on
the Cray T3D is very fine grained, allowing one memory word at a time
to be locked. In practice the {\em atomic update} directive is very
efficient. The presence of this directive, for which there is no
analogue in HPF for example, is the crucial feature which enables this
section of the code to be taken from the serial version and
parallelized with relatively little change. An explicit message
passing code would require a significant rethinking of the strategy for
smoothing particles onto the mesh.
Using this locking approach means that the particles do not need to be
pre-sorted in physical space before the grid can be loaded and no halo
regions are required so there is no final synchronisation phase in
which these regions are summed. The lack of sorting makes the load balance
very good because there are exactly the same number of particles on
every processor (assuming that $N_{\rm pes}$ divides $N$). 

\begin{figure*}
 \centering
 \epsfysize=7cm\epsfbox{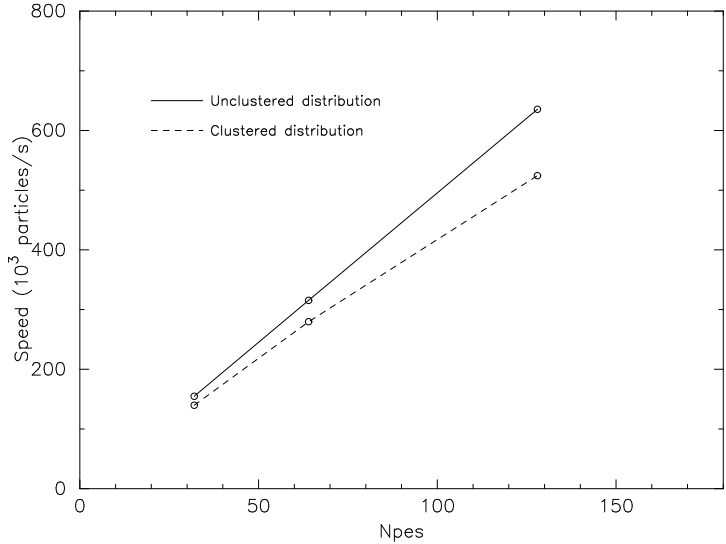}
 \caption{PM speed in thousands of particles per second for
both clustered and unclustered positions. The raw speed is very high
and is almost unaffected by the degree of clustering. The scaling
properties are also 
excellent. The data shown is for a test with $2\times128^3$ particles.}
 \label{pmscaling}
\end{figure*}

Parallel multi-dimensional FFTs are now readily available for the T3D,
but this was not the case when we began to use the machine.  We built
our parallel 3-dimensional FFT in the standard way by performing a
series of short, length $L$, 1-dimensional FFTs in parallel. As the
grid to be transformed is arranged in sheets, two dimensions can be
transformed entirely locally and require no off-processor
communication. In the third dimension a small local array is loaded with
the off-processor values, transformed and then the transformed values
are written back off-processor.

The convolution involves multiplying the Fourier transformed density
field by a Greens function generated from the appropriate force law
(see Figures~1 \& 2).  It is faster to store these Greens functions on
disk than to generate them afresh each time they are needed.  Each
sheet of the convolution is independent of the others  (in fact each
grid point operation is independent) and so the
convolution routine can be parallelised by doing the calculation one
sheet at a time. The potential energy of the calculation is
accumulated as a simple sum over the transformed grid values.

After inverse transformation the grid potentials are differenced and
the forces interpolated back onto the particles. Interpolation is
automatically load balanced because the same number of particles
reside on each node. No blocking problems occur because we are only
reading information about the force values at the nearest grid points,
not writing to them and so no locking is required. One problem is that
as the simulation proceeds the communication overhead grows because
particles will have migrated into another processor's
grid space and so all the force-value requests are off-processor. This
is not a problem for CRAFT as within shared arrays local and remote
accesses take approximately the same time.

\subsection{Building lists}

Neighbour lists are required so that the short range force calculation
can be done efficiently. A tally of how many particles there are in
each PP cell is also required by the refinement placing algorithm.
Building a linked list in parallel is difficult because two processors
might simultaneously try to update the list for the same box. This is
prevented by using an array of locks, one lock for each PP cell so
that only one processor can operate on the part of the list pertaining
to a particular cell at any one time. Arrays of locks are an
undocumented but useful feature of CRAFT.

\subsection{Short-range force}

The short range forces come in two parts, the direct summation over
nearby particles for gravity and the accumulation of the gas forces. 

The short range PP part of the gravity calculation is parallelised by
getting each processor to do the calculation one PP cell at a time.
Before the processing of a cell begins, temporary arrays are loaded
with the properties of the particles in the 26 surrounding PP cells
creating small, entirely local work arrays. This scheme ignores data
locality, (in general the particles being considered will have their
properties stored elsewhere), but given that remote accesses are
limited to one read and one write per particle this scheme works well
in practice. Once a processor has finished a particular cell it finds
out which one to do next by examining a shared counter.

This ``first come, first served'' approach is used to improve load
balancing. As Figure~5 shows most of the 1.6 million or so PP cells
take much less than a second to complete but a few can take much
longer. The very small number of cells taking over 10 seconds all lie
at the edge of a refinement and involve neighbouring cells which lie
in an abutting refinement. As described earlier, it is only pairs of
cells which both lie in the same refinement which benefit from a finer
mesh and hence a reduced PP search length. The construction of a good adaptive
refinement-placing algorithm is difficult and these expensive cells
illustrate instances when the routine should have perhaps placed a
single larger refinement to cover these expensive cell-pairs
(although, in principle, this may not always be possible given the
constraints imposed). It is worth noting that the impact of a
``mistake'' in placing a refinement in terms of relative speed
reduction is far greater when 
the hydra algorithm is executing in parallel than in serial execution
because of the large number of processors which will idle waiting for
the expensive cells to finish. Further, the present algorithm
minimizes the total work assuming serial execution, and does not take
account of the fact that, e.g., 30, ten second refinements may be
preferable to one, 100 second refinement in parallel execution on a
large number of processors. We are
working on an improved refinement-placing algorithm which under nearly
all circumstances removes the occurrence of these very expensive
cell-pairs. By scattering the work in such a fine grained manner a
better load balancing is achieved than if the volume was divided
amongst the processors in sheets or blocks because the difficult cells
tend to be grouped together. Figure~6 details the load balance for the
particle distribution used to produce Figure~5.

The slope of the curve in figure~5 is roughly $t^{-1}$ which indicates
that each decade of refinement-completion times contributes roughly
equal amounts to the cumulative time. The fact that the slope is not
shallower than this is evident in the excellent general load balance
shown in figure~6. On the other hand it is clear that the few
expensive cells lie just on or above extrapolation of the $t^{-1}$
slope in this case, and are thus in danger of dominating the overall
wall clock time, as is clear from figure~6.

\begin{figure*}
 \centering
 \epsfysize=7cm\epsfbox{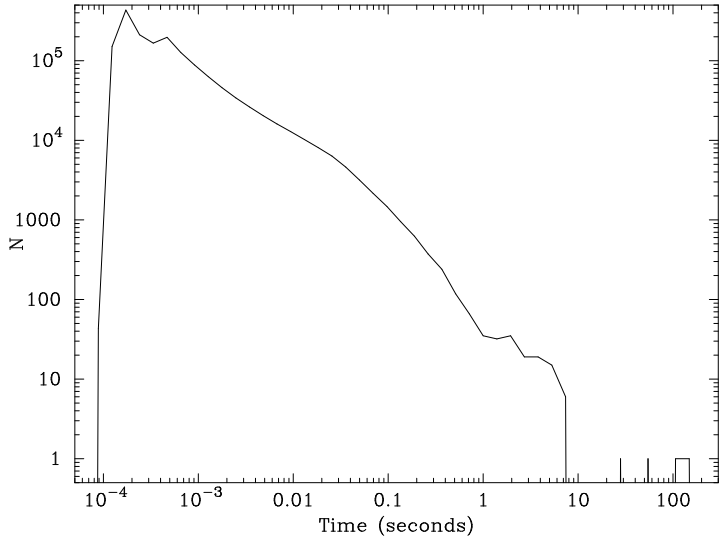}
 \caption{Timing distribution for the $116^3$ PP cells within
a $2 \times 128^3$ particles combined gas and dark matter simulation
at a redshift of 0.5. This distribution of particles is the one used
for the timing breakdown shown in table~1}
 \label{ppcell}
\end{figure*}

\begin{figure*}
 \centering
 \epsfysize=7cm\epsfbox{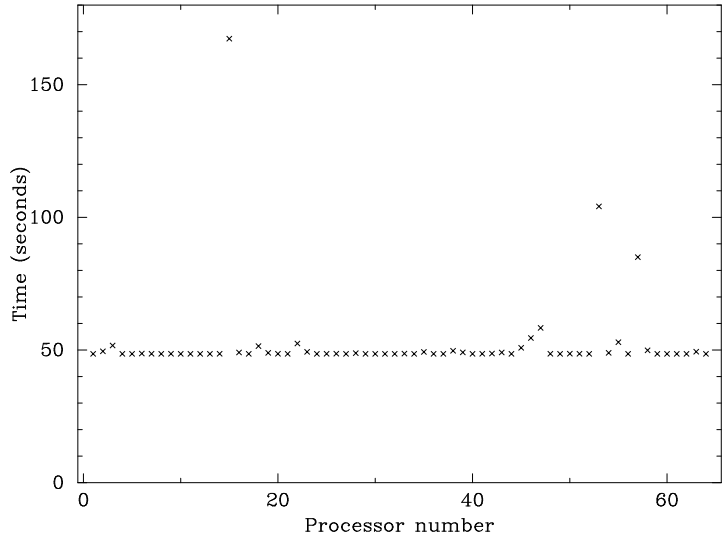}
 \caption{Load balance from the clustered position shown in 
figure~5, the small number of PP cells taking a long time lead to a
large load imbalance.}
 \label{pptime}
\end{figure*}

Parallelising the SPH part of the calculation is not very difficult
because the SPH formalism sits completely within the PP part of the
algorithm. This automatically returns a nearest neighbour list which
is used by SPH to calculate the local density and density gradient
from which the gas forces can be derived.

\subsection{Placing refinements}

Placing refinements in parallel is difficult because they may in
general have any size and distribution, subject to the constraints
that they are cubic and disjoint. These constraints force the
problem to have a non-local character which makes the problem hard to
distribute. Any blocking strategy employed to divide up the
computational volume between the processors leads to boundary
problems: placement of a new refinement is influenced by the
distribution of those already placed.

Refinements are placed using the following strategy. The
numbers of particles in each of 
the PP cells is stored into an array. (The particle number gives a
good measure 
of the PP work). Peaks in this field are located by getting each
processor to search a section of the array for values above a
threshold (typically 40--50 particles per cell). Once a cell
containing more than this threshold is found, the processor walks
uphill through the array until it finds a local maximum in the array.
This walk may take it off-processor, but these segments of the array
are, of course, globally visible under CRAFT. Each new peak is tagged
with a unique number by using a shared counter visible to all the
processors.

Once the positions of the peaks have been found all subsequent
operations are carried out on a single processor as detailed in
section 2.3 above, but as these involve at most a few thousand
refinements this procedure doesn't incur a significant overhead.

The top-level refinement distribution is calculated only every 10
steps because these refinement positions do not change rapidly.
Subsequent levels of refinement are placed every step.  With this
restriction the time taken to place refinements is less than 2 percent
of the total even in difficult highly clustered positions when the
refinement placing algorithm is most expensive.

\subsection{Parallel refinements}

In calculating forces from each refinement we have to perform
essentially the same sequence of operations as are carried out at the
top-level.  The Fourier transform convolution is now no longer
periodic, but this is a small complication as the parallel isolated
FFTs and convolutions are very similar to the wrapped ones.  A further
concern is that the PM grid size and the number of particles involved
are no longer known {\em a priori}.

\subsection{Load balancing}

Load balancing is crucial for an efficient parallel code. 
In general, there are two main types of load balancing,
fine and coarse grained but Hydra also has an extra type of load
balancing that is implicit in the serial method.  We try to balance
the amount of work spent in the long-range force calculation (PM)
against the work spent on the short-range force calculation (PP) by
placing refinements. Placing a refinement increases the PM work and
decreases the PP work, whilst aiming to decrease the total work.

Fine-grained load balancing usually takes place at the very bottom
level of the code and refers to the way in which small units of work
are distributed amongst the processors.  For it to work efficiently a
large number of ``work units'' are required ($N_{\rm unit} >\!> N_{\rm pes}$).
We use this type of load balancing in both the
PM and the PP sections of the code.  In the PM routine there are two
units of work, a particle or a grid point. An equal number of both of
these units are distributed to each processor which efficiently
balances the work because the amount of work associated with each unit
is the same.  In the PP routine the unit of work is a PP cell. These
are distributed to the processors on a `first come, first served'
basis by using a shared counter. The load balance of the PP 
algorithm for a typical clustered step is shown in figure~6.
As noted previously it is hard to achieve a very good load balance 
because of the small number of outlying PP cells with heavy workload.
At this point the load imbalance is a factor of 3, a
typical value for the larger runs under heavy clustering.

Coarse-grained load balancing usually takes place at a much higher
level. Large sections of work are distributed to each processor,
perhaps many subroutines at once.  Within Hydra we use this type of
load balancing for small refinements because it is much more efficient
to do these entirely on a single processor. Once a series of temporary
arrays are loaded all accesses are local.

The overall strategy may be summarised as follows. We complete the top level
in parallel using the whole machine, then we complete all refinements
above a certain size, again using the whole machine and then we employ
a task farm approach to complete the small refinements one to a
processor.  Typically there are 10 big refinements and 1000 small
refinements for 4 million particles. Figure~7 shows the load balance
for the task farm employed to do the first level refinements on a
typical step. Around one quarter to one third of the cpu time is
wasted by processors standing idle at this stage.

\begin{figure*}
 \centering
 \epsfysize=7cm\epsfbox{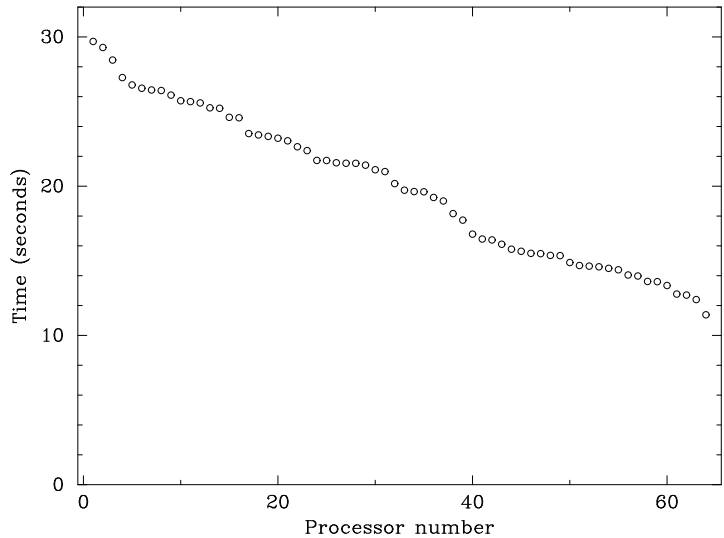}
 \caption{Distribution of total times per processor spent calculating
task-farm refinements at the first level.
The distribution of refinements is the same as that used to
produce figures~5 \& 6 and table~1}
 \label{timeref}
\end{figure*}

\section{Performance}

\subsection{Testing} 

One of the major advantages of using CRAFT is that the parallel code
is immediately portable to a workstation. This considerably eased
development by allowing the output to be tested against the original
code, tests of which have been published (Couchman, Thomas \& Pearce
1995).  Once the parallel code was completed we made 3 test runs, each
a complete simulation. Two used the new code, on a workstation and the
T3D and the third the original serial code.  All three runs produced
results which were the same up to the accuracy expected due to
different execution order. All algorithm development and
testing takes place on workstations using the production parallel
code.

\subsection{Speed}

For complex codes, and in particular adaptive codes, the operation
count or Mflop rate is not a good measure of efficiency because the
programmer's art lies in reducing the total operation count required
to perform a particular task. This often involves increased complexity
at the expense of the Mflop rate even though the required cpu time has
been reduced. In practice, however, the Mflop rate is a widely used
measure of performance and so we present the results here. Under CRAFT
Hydra achieves $12\,$Mflops per processor on 256 processors of the
T3D. Running a small version of Hydra on a single processor, and hence
not under CRAFT, the code achieves a Mflop rate which is nearly 3
times higher. This gives a clear indication of the penalty that is
being paid for parallel execution in this implementation of the
code. Part of this performance loss is simply because of the nature of
the algorithm and the inevitable inefficiencies that occur with
parallel programming, a part is the penalty that is incurred by using
CRAFT which, by hiding a significant amount of communication and data
management, has eased the programming task. In what follows we have
chosen to measure code speed in terms of the number of particles which
can be computed per second as this, combined with the number of steps
required, is ultimately what determines the size of problem that can
be addressed and the time in which it will complete.

\begin{figure*}
 \centering
 \epsfysize=7cm\epsfbox{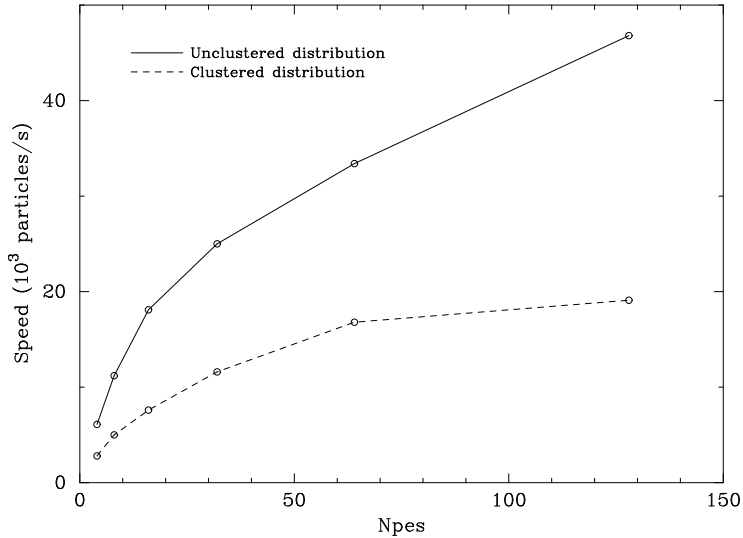}
 \caption{Code speed in thousands of particles per second as a
function of number of processors for
both smooth and clustered positions for a small $2 \times 64^3$ particle
combined gas and dark matter cosmological simulation}
 \label{speed}
\end{figure*}

Hydra is extremely fast when the particles are unclustered.  As
structure forms the mean number of neighbours grows, increasing the
work significantly. The amount of communication required for the
parallel code also rises sharply because under CRAFT the PP
calculation is expensive.  The speed and scaling properties of the
full code are shown in Figure~8. We plot the number of particles which
can be computed per second on various numbers of processors for a
small combined gas plus dark matter simulation which contained $64^3$
particles of each species. A problem of this size just fits on 4 T3D
processors but, as shown in the figure, is too small to scale well
above 32 processors. The plot gives timings both for the initial
smooth distribution and for a highly evolved position at a redshift of
0.5. Rates for larger gas-plus-dark matter runs with $2\times 128^3$
particles on 256 processors have been measured at $6.6\times 10^4$
particles per second unclustered, dropping to $1.1\times 10^4$
particles per second under very heavy clustering. In runs with only
dark matter with $256^3$ particles we have achieved peak rates of
$3.3\times10^5$ particles per second on 256 processors unclustered,
and, under moderate clustering, $6.8\times10^4$ particles per second
on 128 processors and 
$8.4\times10^4$ particles per second on 256 processors.

\subsection{Load Balance}

Initially the load balance is very good because there is very little
PP work and the PM section is well balanced. Once clustering develops
we encounter three problems. Firstly, some of the PP cells take a long
time to complete and even with the very fine grained balancing
employed by the PP routine there is an imbalance while one processor
finishes a particularly difficult cell (see Figure~6). Secondly, the
large refinements that are done in parallel across the full machine
are really too small to be done in this way (but too big for a single
processor). The maximum number of processors 
that should be used for the FFT routine is $L_{\rm mesh}$, the size of the
PM grid. In practice, $L_{\rm mesh}=64$ for these refinements and the
time taken does not scale well if more than 32 processors are
used. CRAFT does not have a facility that would allow the assignment
of a subset of the processors to a task, being limited to either using
all the processors or just one processor. Finally, a refinement cannot be
started before its parent has been completed. This means that the
small refinements cannot all be completed as a single task farm but must
be done one level at a time. Unfortunately, the time then scales as
the sum of the times of the most expensive refinement at each
level. As shown in figure~7 this leads to a 
load imbalance while one processor finishes the most difficult
refinement at each level. 

Table~1 gives the relative timings for the
main sections of the code on a typical clustered step within a $2
\times 128^3$ particle combined gas and dark matter cosmological
simulation.  Over half of the time is spent in the PP routine at the
top-level.  There are four levels of refinement. The single
fourth-level refinement takes longer than the 17 third level ones
combined 
because 64 processors are available so these all execute in
parallel. The biggest saving in terms of load balance could be
achieved in the PP routine where perhaps 100 seconds of the nearly
300 second step could be saved if it was possible to either split up
individual PP cells over several processors or to place the
refinements so that these cells are avoided.

\begin{table*}
 \begin{tabular}{lclcc}
Routine		  &      & Subroutine && Number\\
Top level	  & 66.8 & && 1\\
		  &      & \PP & 55.9 &\\
		  &      & \PM & 6.78 &\\
		  &      & list sort & 3.30 &\\
		  &      & placing refinements & 0.81 &\\
Big refinements   & 14.0 & && 3\\
		  &      & 1 & 3.77 &\\
		  &      & 2 & 5.06 &\\
		  &      & 3 & 5.17 &\\
Small refinements & 19.2 & &\\
		  &      & level 1 & 12.6 & 358\\
		  &      & level 2 & 4.66 & 182\\
		  &      & level 3 & 0.83 & 17\\
		  &      & level 4 & 1.08 & 1\\
\end{tabular}

 \caption{Percentage time required by different coarse grained sections of the
code, taken from redshift 0.5 in a $2 \times 128^3$ particle combined gas
and dark matter run}
 \label{timediag}
\end{table*}

\subsection{Memory requirements}

The number of words of storage required by Hydra is,
\begin{eqnarray}
N_{\rm words} &=& 2(L_{\rm max}^3+2L_{\rm pmax}^3 + 2N_{\rm pes}L_{\rm rmax})+ \\ \nonumber
   & &   19(N_{\rm max} + N_{\rm pmax} +N_{\rm pes}N_{\rm rmax})+
5N_{\rm pes}n_{\rm max}, 
\end{eqnarray}
where, $L_{\rm max}$ is the maximum size of the top level PM mesh,
$L_{\rm pmax}$ is the maximum size of the parallel refinement meshes,
$L_{\rm rmax}$ is the maximum size of the small (task-farmed)
refinement meshes, 
$N_{\rm pes}$ is the number of processors, $N_{\rm max}$ is the maximum number
of particles, $N_{\rm pmax}$ is the maximum number of particles within a
parallel refinement, $N_{\rm rmax}$ is the maximum number of particles
within a small refinement and $n_{\rm max}$ is the maximum number of PP
neighbours allowed.

\begin{figure*}
 \centering
 \epsfysize=7cm\epsfbox{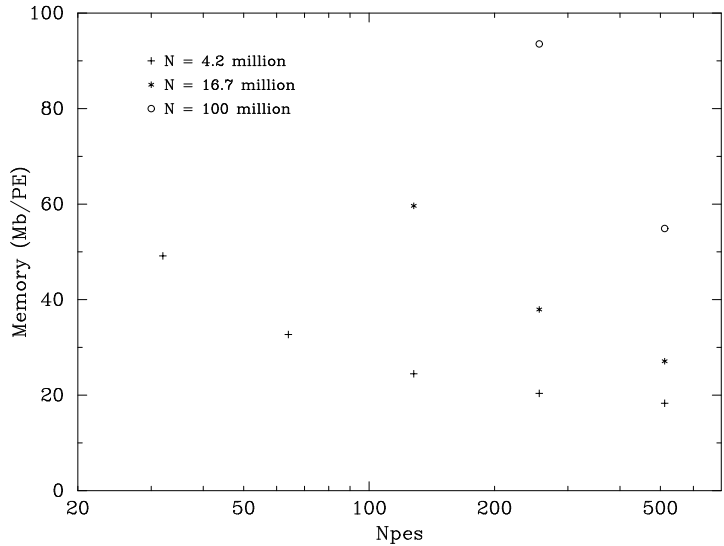}
 \caption{Memory usage of parallel Hydra for several different values
of $N$. We are capable of running a 100 million
particle gas and dark matter run on a 512 processor T3D.}
 \label{mem}
\end{figure*}

Typically we set, $L_{\rm max} = 2N_{\rm max}^{1 \over 3}$, $N_{\rm
pmax}=128^3$, $L_{\rm pmax}=128$, $N_{\rm rmax}=32^3$, $L_{\rm
rmax}=32$ and $n_{\rm max} = 64^3/2$. Figure~9 shows a plot of the
memory requirements of the code for various values of $N$ against
number of processors.  There are $64\,$ Mb of memory available per
processor on the T3D in Edinburgh.  Equation 2 is a good measure of
the memory usage because the 16.8 million particle simulation does in
fact run on 128 processors.

These figures are for the full SPH code. Without gas the storage
requirement can be reduced substantially. In this case the number of
arrays needed to store the particle properties can be reduced from
19 to 13 trivially and even further to 7 if the timestepping scheme is
modified.

\section{Summary}

A parallel adaptive grid code based on Couchman's (1991) algorithm has
been implemented on the Cray T3D using CRAFT. We have incorporated SPH
following the prescription of Couchman, Thomas \& Pearce (1995). The
code, Hydra, is capable of running $2 \times 64^3$ particle combined
cooling gas and dark matter cosmological simulations in $\sim 600$
processor hours.  These runs are very useful for exploring parameter
space in order to target larger $2 \times 128^3$ particle runs. We
have completed a sequence of these, the results of which will be
presented elsewhere (e.g., Pearce \etal 1997).

The parallel version of Hydra can also be used for more traditional
dark matter only simulations. We have completed a series of $256^3$
(16.8 million) particle models in a variety of cosmologies (Thomas
\etal 1997, Jenkins \etal 1997). Parallel Hydra was also used as part
of the recent cluster comparison exercise (Frenk \etal 1997) where
many different hydrodynamic codes where used to perform the same
simulation.

\section*{Acknowledgments}

We would like to thank the staff of the Edinburgh Parallel
Computing Centre for their help throughout this project.
FRP was an EPSRC PDRA working for the Virgo Consortium. 
HMPC is supported by NSERC of Canada.
We acknowledge a NATO (CRG 920182) travel grant which facilitated our
interaction.

\section*{References}

\noindent Aarseth, S. J., 1963, \MN, 126, 223

\noindent Appel, A., 1985, SIAM J. Sci. Stat. Comp., 6, 85

\noindent Barnes, J. E., Hut, P., 1986, \Nat, 324, 446

\noindent Cole, S., Lacey, C., 1996, astro-ph/9510147

\noindent Couchman, H. M. P., 1997, In Proceedings of the 12th Kingston Meeting on Theoretical Astrophysics, eds. Clarke, D. and West, M.

\noindent Couchman, H. M. P., 1991, \ApJL, 368, L23

\noindent Couchman, H. M. P., Thomas, P. A., Pearce, F. R., 1995,
\ApJ, 452, 797

\noindent Dav\'e, R., Dubinski, J., Hernquist, L., 1997, astro-ph/9701113

\noindent Dubinski, J., 1996, NewA, 1, 133

\noindent Efstathiou, G., Davis, M., Frenk, C. S., White, S. D. M.,
1985, \ApJS, 57, 241

\noindent Efstathiou, G., Frenk, C. S., White, S. D. M., Davis, M.,
1988, \MN, 235, 715

\noindent Efstathiou, G., Eastwood, J. W., 1981, \MN, 194, 503

\noindent Evrard, A. E., 1990, \ApJ, 363, 349

\noindent Ferrell, R., Bertschinger, E., 1994, Int. Jour. Mod. Phys.
C, 5, 933

\noindent Ferrell, R., Bertschinger, E., 1995, In Proceedings of the
1995 Society for Computer Simulation Multiconference

\noindent Frenk, C. S. \etal, 1997, \prep

\noindent Greengard, L., Rokhlin, V., 1987, \JCP, 73, 325

\noindent Hockney, R. W., Eastwood, J. W., 1981, Computer Simulation
Using Particles, McGraw-Hill

\noindent Jenkins, A. R., Pearce, F. R., Thomas, P. A., Frenk, C. S.,
Couchman, H. M. P., White, S. D. M., Colberg, J. M., Hutchings, R.,
Peacock, J. A., Efstathiou, G. P., Nelson, A. H., 1997, \prep

\noindent Jernigan, J. G., 1985, In IAU Symposium No.127, ed. J. Goodman,
P. Hut, pg 275, Dordrecht, Reidel

\noindent Miller, R. H., 1978, \ApJ, 223, 122

\noindent Miller, R. H., Smith, B. F., 1980, \ApJ, 235, 421

\noindent Monaghan, J. J., 1992, \ARAA, 30, 543

\noindent Pearce, F. R., Thomas, P. A., Couchman, H. M. P., 1993,
\MN, 264, 497

\noindent Pearce, F. R., Thomas, P. A., Couchman, H. M. P., 1994,
\MN, 268, 953

\noindent Pearce, F. R., Couchman, H. M. P., Jenkins, A. R., Thomas,
P. A., 1995, in Dynamic Load Balancing on MPP Systems

\noindent Pearce, F. R., Thomas, P. A., Jenkins, A. R., Frenk, C. S.,
Couchman, H. M. P., White, S. D. M., Colberg, J. M., Hutchings, R.,
Peacock, J. A., Efstathiou, G. P., Nelson, A. H., 1997, \prep

\noindent Salmon, J., 1991, \PhD, California Institute of Technology

\noindent Sellwood, J. A., 1987, \ARAA, 25, 251

\noindent Steinmetz, M., 1996, \MN, 278, 1005

\noindent Thomas, P. A., Couchman, H. M. P., 1992, \MN, 257, 11

\noindent Thomas, P. A., Pearce, F. R., Jenkins, A. R., Frenk, C. S.,
Couchman, H. M. P., White, S. D. M., Colberg, J. M., Hutchings, R.,
Peacock, J. A., Efstathiou, G. P., Nelson, A. H., 1997, \prep

\noindent van Albada, T. S., van Gorkum, J. H., 1977, \AaA, 54, 121

\noindent von Hoerner, S., 1960, Zeit. Ap., 50, 184

\noindent Villumsen, J. V., 1982, \MN, 199, 493

\end{document}